# Dementia in England: Quantifying and analysing modifiable risk

**Christopher Drowley & Luke Burns**

University of Leeds, Leeds, United Kingdom

# Abstract

The prevalence of dementia is set to explode throughout the 21$^{st}$ century. This trend has already started in developed countries and will continue to place heavy pressures on both public health and social care services across the world. No cure for dementia is likely within the foreseeable future, however, medical research highlights the potential to diminish the risk of dementia onset. Over one-third of dementia cases may be preventable if certain risk factors are addressed at the individual, clinical, and population level. This research further explores these modifiable risk factors and quantifies areal risk through the use of a composite index. The index operates at National Health Service Clinical Commission Group level to assess spatial differences across England. Clear spatial patterns are observed between the north and south of the country, and between London and the remainder of the country. The framework adopted in this research provides a firm foundation upon which similar indices could be produced, potentially at finer spatial resolutions, incorporating more informed local knowledge and data on relevant dementia risk factors.

# Keywords

# Background and rationale

Advances in public health and medicine have considerably improved health globally, allowing people to live longer and healthier lives (GBD, 2016). Conversely, the prevalence of harmful chronic diseases continues to grow. In recent years, dementia has been acknowledged as potentially the "*greatest global challenge for health and social care in the 21$^{st}$ century*" (Livingston et al, 2017, p.2673). Dementia refers to a set of symptoms related to the gradual decline of the mental cognition, which often impairs motor functions, memory and mood. This leaves those with the condition with a decreased capacity to perform everyday tasks and eventually deprives their ability to care for themselves (NHS, 2017). The word itself derives from the Latin 'de' meaning 'to depart', and 'mens' meaning 'mind', and serves to reflect the impact the condition has on people in that they lose the character and capabilities they once had. In a recent survey of adults over the age if 50, dementia was found to be the most feared health condition, and this fear increased with age indicating that exposure to the disease allows for a fuller comprehension of the insidious impact the symptoms bring (SAGA, 2018).

The Alzheimer's Society, a UK dementia care and research charity, conducted a large-scale report in 2007 which exposed the staggering financial and human impact currently wrought by dementia. The report found that 1 in every 88 people in the UK were living with dementia and estimated future prevalence trends, primarily driven by an aging population, of increases of 38% over the next 15 years and 154% over the next 45 years, the scale of which can be observed in Figure 1 (Alzheimer's Society, 2007).

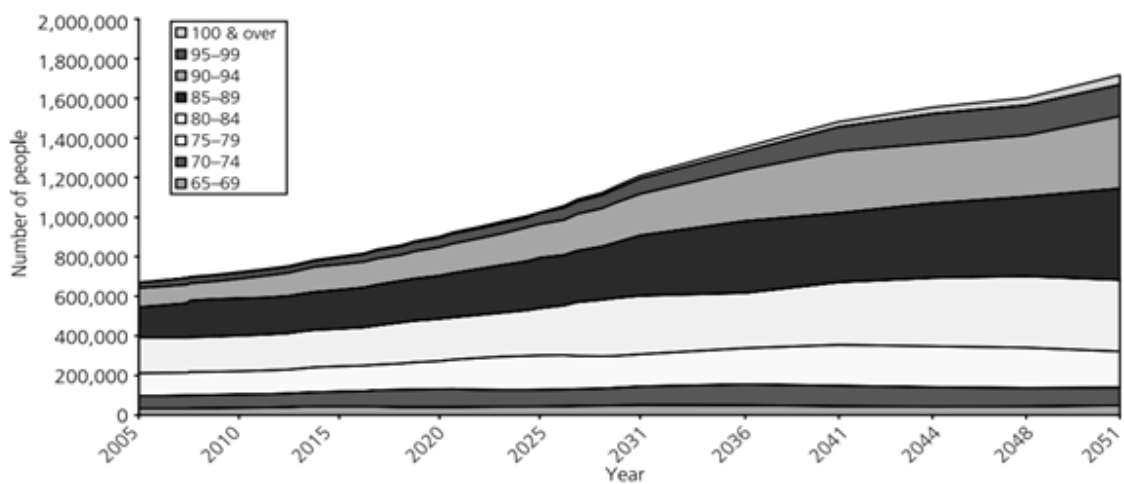

*Figure 1: Projected number of people with late onset dementia by age group (Alzheimer's Society, 2007)*

One of the report's most tangible results was the production of UK prevalence estimates, which enables health authorities to work towards more definitive targets in terms of dementia diagnosis. As a slowly debilitating condition, with societal and personal stigma, and a difficulty in identification, rates of diagnosis were substantially lower in 2007 compared with present day (NHS Digital, 2017).

In 2014, the Alzheimer's Society updated their report and renewed the original Delphi consensus estimates, providing more precise prevalence data nationwide (Alzheimer's Society, 2014). Based primarily on age-related risk derived from the English Longitudinal Study of Aging, these estimates have a variety of problems. Such problems include; the small catchment areas sampled, numerous selection biases of only those who received an early-onset dementia diagnosis, limitations of prevalence in differing care settings, in addition to ignoring other dementia risk factors (Alzheimer's Society, 2014). While these prevalence estimates are crucial for use in a range of public sector settings, it is unknown how these relate to future spatial dementia patterns or risk. Moreover, as there is no expectation of a cure in the short term (Norton et al, 2014), focus must shift elsewhere.

## Past research: Dementia modifiable risk factors

There is an increasing body of literature on the potentially modifiable risk factors of dementia, elimination of which would delay the onset or even prevent certain forms of the disease materialising (Livingston et al, 2017). Research continues to uncover the broad aspects of dementia risk, yet one clearly identifiable group of risk factors revolves around the concept of '*what is good for the heart is good for the head*', a common belief which drives current GP guidelines around dementia prevention in the UK (NICE, 2013). In a recent commission on dementia conducted by the Lancet, Livingston et al (2017) created a model of dementia risk which proposed that 9 potentially modifiable risk factors account for up to 35% of dementia cases (see Figure 2). Targeting these risk factors should dominate current efforts in tackling dementia, and any action is likely to reduce future human and financial costs associated with the condition (Schwarzinger et al, 2018).

The geography of dementia prevalence and incidence has been well researched and reported in the UK, however, whether areas display spatial variations in terms of modifiable risk factors has not been investigated to date. Furthermore, given that many modifiable risk factors display associations with deprivation, there is a surprising lack of literature linking deprivation to dementia, particularly from a spatial analysis perspective. Thus, it can be argued that a better understanding of areal modifiable dementia risk and the links to deprivation is needed, and this, in part, provides the research rationale for this study.

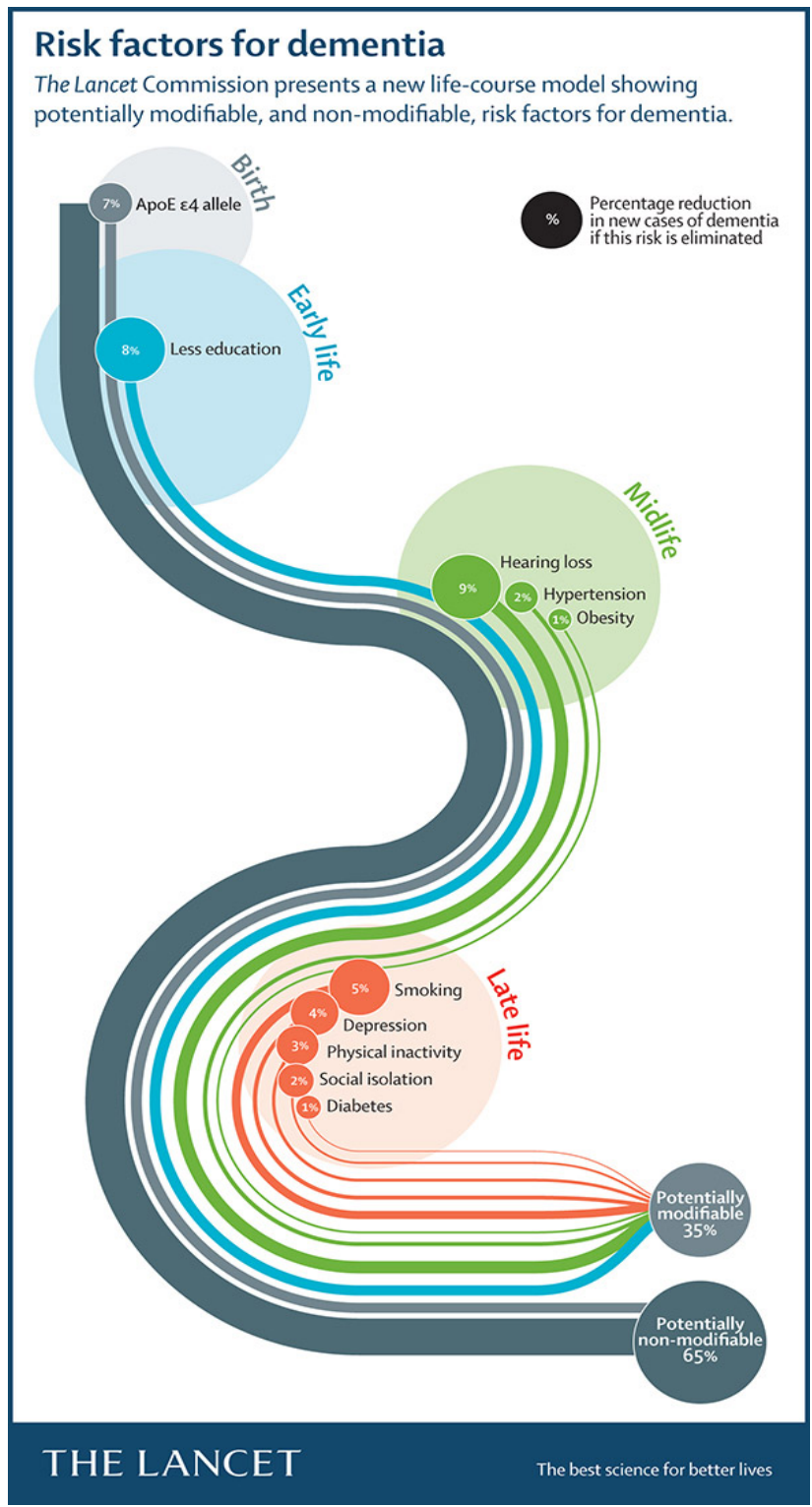

*Figure 2: Contribution of modifiable risk factors to dementia (Livingston et al, 2017)*

# Dementia and dementia risk

Dementia is classified as a neurocognitive disorder with varying degrees of severity by the Diagnostic and Statistical Manual of Mental Disorders, Fifth Edition (Alzheimer's Society, 2017). Moreover, dementia is a general term which covers a range of progressive organic brain diseases symbolised by problems of short-term memory and other cognitive shortfalls (Holmes and Amin, 2016). The direct causes of dementia most commonly include; Alzheimer's disease (~62% of cases) and Vascular dementia (~17%), with smaller proportions attributable to Lewy bodies (~4%), Parkinson's disease (~2%), frontotemporal disorder (~2%) and other factors combined (~3%) – an important facet of dementia diagnosis is that mixed pathologies are not uncommon (Chapman, 2006). It is important, where possible, to determine the direct cause, as differing forms of dementia present different courses, patterns of symptoms, and responses to treatment (Burns and Iliffe, 2009). There is some debate as to the current approach of dealing with dementia in society in that dementia is an umbrella term for much of the above. Haeursermann (2017) calls for a disaggregation of the syndrome and argues that the umbrella term is unhelpful and masks understanding and hampers potential solutions. However, as dementia is heterogeneous and risk factors can vary, and may also co-exist (Livingston et al, 2017), dealing with dementia in this aggregate form still represents a useful angle on which to base this research.

Literature on dementia risk is understandably concentrated within the fields of medicine, epidemiology, and psychiatry. Dementia becomes increasingly prevalent as one ages and this is understandably acknowledged as the primary risk factor (Holmes and Amin, 2016; Wu et al, 2016). However, this is at least in part due to the accumulative risk over the life course, and dementia need not necessarily be a consequence or inevitably of aging (Livingston et al, 2017).

Figure 3 illustrates the increasing prevalence of dementia in society and shows how dementia risk begins to increase year-on-year beyond the age of 65 and continues to escalate until (on average) 80-84 for males and 85-89 for females (Prince et al, 2014). Additionally, dementia prevalence in women is generally higher than men and doubles with every five-year increment in age (Van der Flier and Scheltens, 2005). Besides increased risk, dementia presents additional challenges for females as the majority of the care burden has been placed on women (Prince et al, 2014). Evidence for familial risk owing to genetic factors is strong, however, this accounts for under 7% of Alzheimer's-related dementia (Livingston et al, 2017; Holmes and Amin, 2016).

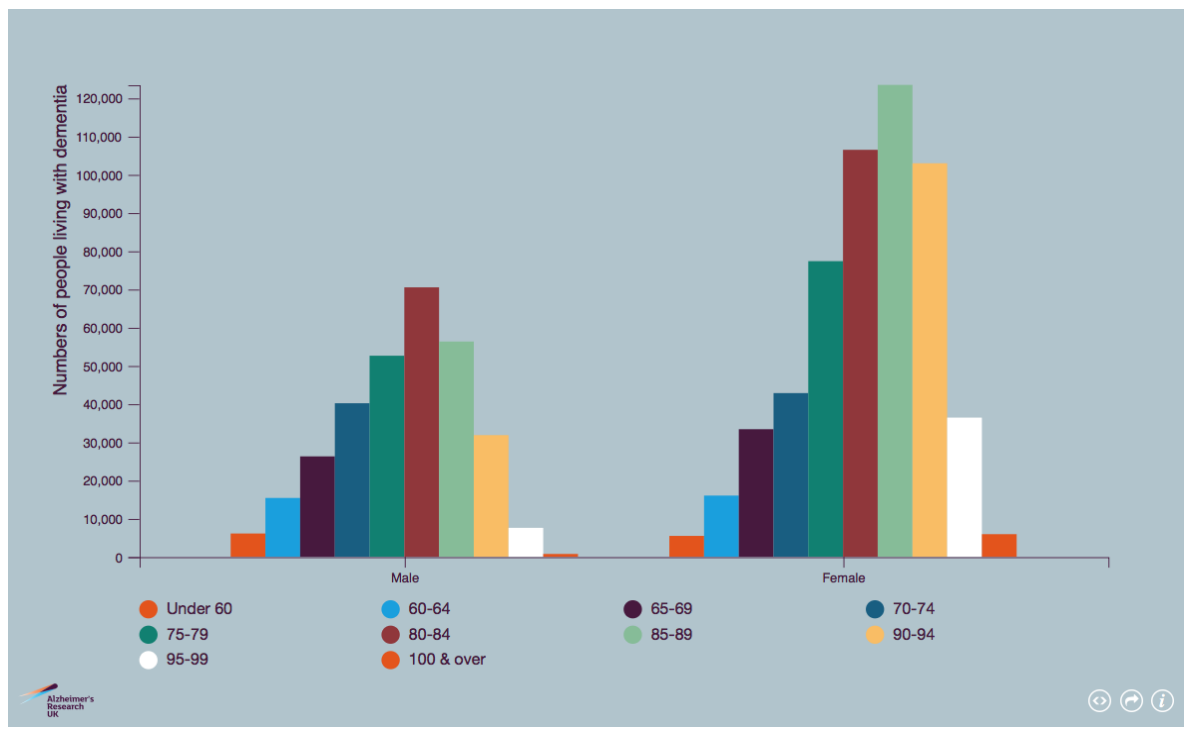

*Figure 3: Age breakdown of no. people living with dementia in 2015 (Prince et al, 2014)*

Previous work has identified the potential for modifying dementia risk (Dixon, 2006; Stern, 2005) and research within this domain was stimulated following the 2013 Group of Eight (G8) meeting, which oversaw the creation of the World Dementia Council to lead in the effort to fight dementia. A key element of the council's work remains an action plan to transform knowledge on dementia risk (Baumgart et al, 2015). The G8 commitment represented a step-change in preventative dementia research, with evidence for the multi-faceted nature of dementia risk having been generated, reviewed and condensed through large scale systematic reviews, meta-analyses and longitudinal studies, all of which present strong evidence for a multitude of 'potentially' modifiable risk factors (Saczynski et al, 2010; Lin et al, 2011; Byers and Yaffe, 2011; Anstey et al, 2011; Cheng et al, 2012; Mathews et al, 2013; Shankar et al, 2013; Strand et al, 2014; Lincoln et al, 2014; Norton et al, 2014; Alzheimer's & Dementia, 2015). If these risks were targeted using effective modifications in behaviour, specific medical diagnoses and local interventions or population-level initiatives, there is a potential for a reduction of up to 35% of new dementia cases across the world (Livingston et al, 2017).

The aforementioned efforts have provided a wealth of information, with early advice primarily focused on risk factors which, if controlled throughout mid-life (aged 45-64), present huge promise in reducing individual dementia risk. Whilst this body of evidence was uncovered through thorough research and consultation, it is led by the long-time common-sense approach of "*what's good for the heart is good for the head*" (Brody, 2005). After reviewing the published guidelines from the National Institute for Health and Care Excellence, these guidelines highlight similar mid-life approaches to delaying or preventing the onset of dementia, disability and/or frailty in later life (NICE, 2013). The following dementia risk factors are identified; smoking, exercise, reducing alcohol consumption, improving diet and maintenance of a healthy weight. Targeting mid-life risk began the process and provided a suitable platform for understanding dementia risk, arguably accelerating preventative efforts (Livingston et al, 2017).

The contribution various genetic, environmental and lifestyle factors have on dementia onset has been increasingly scrutinised over the past 20 years. In an effort to collate this vast body of research, the Lancet commission report on Dementia Prevention, Intervention, and Care conducted by Livingston et al (2017), provides a contemporary evidence-based resource as a contribution to global efforts to improve the lives of people with dementia and their careers, and limit the future impact (Prince, 2017). The review interpreted and assessed knowledge on effective interventions on dementia progression and current best medical practice in caring for those with dementia. Of primary interest is the huge contribution provided in understanding dementia prevention. This is especially relevant given that cures for Alzheimer's disease and other forms of dementia are not forthcoming. With the commission concluding that the best approach is to tackle the problem prior to dementia onset (Prince, 2017; Livingston et al, 2017).

The creed '*prevention is better than cure*' has further underlined the growing interest in modifiable risk factors (Livingston et al, 2017, p.2677). Pursing this interest, the Lancet commission quantified and generated a model of total dementia risk at the population level. The results of this Population Attributable Fraction (PAF) model are evident in Figure 2 where 65% of dementia risk remains potentially unmodifiable (i.e. linked to age and genetics). Advances in knowledge of other risk factors and genetic treatments are likely to increase dementia prevention potential, however, the model already attributes 35% of dementia risk to factors which may be acted upon today to reduce or delay onset of dementia in the future (Livingston et al, 2017).

Livingston et al (2017) built on the contributions of others who had previously reviewed evidence for a multitude of risk factors (including Lafortune et al, 2016; Wu et al, 2016; Clare et al, 2017), however, Livingston and colleagues went further in their epidemiological assessment of the impact of hearing loss as a prime risk factor of dementia. As Figure 2 illustrates, the model continued the life-course process from earlier mid-life risk factors and separates dementia risk into different life stages. For the research conducted herein, these risk factors have been assumed to be relevant throughout the life course.

The estimates within the PAF model portray a theoretical reduction of 35% of new cases over a given period of time if particular risk factors were completely eradicated (Livingston et al, 2017). The 9 potentially modifiable risk factors identified include; midlife hearing-loss, lack of secondary-education, smoking, untreated depression, physical inactivity, high blood-pressure, social isolation, type-2 diabetes, and obesity (Lafortune et al, 2016; Wu et al, 2016; Clare et al, 2017; Prince et al, 2014; Prince, 2017; Baumgart et al, 2015; AS, 2015; Livingston et al, 2017). Whilst Livingson et al (2017) and the aforementioned studies provide a comprehensive resource to understand dementia risk, an independent review of relevant literature was also undertaken in a bid to identify any additional potentially modifiable risk factors. The results are shown in Table 1 and do include overlap with the above-mentioned studies.

| *Modifiable Risk factor* | Source | Summary |
| --- | --- | --- |
| *Obesity* | Rönnemaa et al (2011) | A Swedish longitudinal study followed 2,268 males from age 50, which periodically collected data on blood pressure and samples, BMI, smoking status and so on, and found genetic and vascular risk factors, such as obesity, **smoking,** and **hypertension** increased dementia risk. |
| | Loef and Walach (2013) | By examining a multitude of meta-analyses on obesity as a dementia-risk this study forecast future dementia prevalence trends attributable to obesity, while accounting for the impact of obesity prevalence change in mid-life. They found that in China and the US obesity is set to significantly contribute to increased dementia prevalence. |

|  | Anstey et al (2011) | A meta-analysis of 15 studies evaluated the impact certain body mass index categories have on dementia. The measure allows confident conclusions of an increased risk of dementia for those defined as 'underweight', 'overweight', or 'obese'. |
|---|---|---|
| *Excessive Alcohol Consumption* | Balakrishnan et al (2009) | Found that regular and low-to-moderate consumption of alcohol may infer health benefits to CHD and dementia, however overconsumption is associated with increased risk. |
|  | Harper (2009) & Ridley et al (2013) | Evidence as to the damaging effects of low to moderate alcohol consumption is unclear, often contradictory, and suffers from varying definitions and methodological approaches, however the picture is much clearer when considering excessive and prolonged use, which has been long associated with cognitive decline and more recently direct dementia risk. |
|  | Schwarzinger et al, 2018 | Robust recent evidence by Schwarzinger et al (2018) adds much weight to the contribution of this risk factor, previously identified in Livingston et al (2017) as requiring further investigation, a nationwide French study, included 31,624,156 adults and determines excessive alcohol consumption as the largest modifiable dementia risk factor. |
| *Education* | Strand et al (2014) | Using the death registry to identify dementia related deaths, 90,843 previous participants of the Counties Study and Cohort of Norway study, were followed up. Dementia risk was decreased with those with higher or middle educational levels, whereas risk increased in those with lower levels of education. This association accounted for increased risk from **obesity**, **smoking**, **hypertension**, **diabetes**, and other factors. |
| *Hearing Loss* | Lin et al (2011) & Lin and Albert (2014) | Direct observations of participants in the Baltimore Longitudinal studies provide evidence and rationale for this risk factor, in that a reduction in one's hearing subsequently reduces the amount of auditory processing in the brain, this has been shown to lead to a reduction in brain size similar to the results of non-muscle use. Further, in 2014 a review of various longitudinal studies confirmed the strength of the relationship between hearing loss and dementia. |
| *Smoking* | Anstey et al (2007) | A meta-analysis of 19 studies which found a statistically significant increased risk of Alzheimer's disease and cognitive decline in later life attributable to smoking. |
| | Rusanen et al (2011) | Identified heavy smoking as a major risk factor in mid-life. Participants in a health survey conducted from 1978-1985 were followed up, smoking more than 2 packets of cigarettes a day contributed to 100% increase in dementia risk compared with non-smokers. |
| *Depression* | Saczynski et al (2010) | Using 949 elderly (mean age of 79) Framingham cohort participants, a baseline level of depression was assessed and followed up 17 years later. Increases in depression were associated with a significant increase of dementia prevalence. |
| *Hypertension* | Baumgart et al (2015) | In a comprehensive evaluation of well evidenced dementia risk factors, including **diabetes, obesity, smoking, physical activity, diet, alcohol consumption, social isolation, education, sleep, depression,** and hypertension, Baumgart et al (2015) provide an overview of the weight of evidence effective of treatment of hypertension has on dementia and cognitive decline risk. |
| *Diabetes* | Gudala et al (2013) | A meta-analysis consisting of 28 observational studies showed increases of 56%-127% depending on the dementia sub-type. |

|  | Vagelatos et al (2013) | A systematic review revealed the weight of evidence on the impact of diabetes on Alzheimer's Disease, earlier studies displayed less confidence in this conclusion in contrast to more recent studies. Figure 2.7 graphs the odds ratio for the impact of diabetes on dementia, with studies to the right of the line and smaller whiskers showing increased odds of this relationship and vice versa. |
|---|---|---|
| *Lack of Sleep* | Spira et al (2013) | A cross-sectional study of 70 participants from the Baltimore Longitudinal Study of Aging found an association between reports of shorter sleep and lower sleep quality with greater Aβ-Amyloid burden, which in turn is linked to Alzheimer's disease. |
|  | Xie et al (2013) | Found that sleep may allow for a restorative effort in the brains of mice in the removal of neurotoxic waste within the nervous system. |
| *Physical Activity* | Arasaratnam et al 2016 | Reviewing the vast body of the literature Arasaratnam et al (2016) state "physical activity is a potentially modifiable protective factor with established collateral benefits, such as the prevention of cardiovascular disease and diabetes". |
|  | Wills et al (2012) | By examining Medicare claims from 18,670 participants of the Cooper Centre Longitudinal Study the authors investigated the association between a variety of chronic conditions including dementia and physical activity. With a median follow-up of 26 years differences in prevalence were observed between high and low quantile levels by fitness. |
| *Diet* | Hardman et al (2016) | Identified, by analysing longitudinal and cross-sectional studies, that the consumption of a Mediterranean-style diet of fruits, vegetables, and fish shows encouraging potential in the reduction of dementia when compared to a diet of |

| | | overconsumption of dairy, red meat, and sugars (including from carbohydrates). |
|---|---|---|
| Social Isolation | Pillai and Verghese (2009) | In a review of 8 longitudinal studies, conducted across the world, the role of social networks on dementia risk was assessed. Results suggest that increased levels of social engagement, as well as marriage or co-habitation provided a protective effect towards the development of dementia. |

*Table 1: Potentially modifiable dementia risk factors*

# Developing a spatial index to quantify dementia modifiable risk

This section will detail the procedure followed in order to create a composite index examining and ranking areas in terms of their potential modifiable dementia risk – hence the index is named: The Index of Modifiable Dementia Risk (abbreviated to IMDR). The approach adopted is intended to allow for recreation at a smaller geographic scale, should the data become available in future. A common structure is followed with regards to the generation of the IMDR, following processes set by Lucy and Burns (2017) and the official handbook on constructing composite indicators (OECD, 2008). In summary, these processes include: (1) Defining the purpose of the index, (2) determining the choice of the spatial scale, (3) sourcing the input variables, (4) pre-processing the data (including testing for multicollinearity, undertaking normalisation, determining polarity and assigning variable weights), (5) generation of the index score, and finally (6) visualisation (including mapping and radar diagrams). The index was also validated against external sources using approaches such as correlation analysis to ensure its robustness and credibility.

## (1) Defining the purpose of the index

The purpose of any composite index is to simplify phenomena and provide information when direct measures are unfeasible, in this case relating to dementia risk as qualified by known risk factors. As noted previously, a cure for dementia is unlikely in the short-term and as such focus has turned to prevention and delaying onset (Alzheimer's Society, 2014; Livingston et al, 2017). The theoretical framework underpinning this work is that there are lifestyle changes, population-level initiatives and medical interventions which may impact an individual's risk of developing dementia. In short, this is an index to quantify and benchmark areas by dementia risk based on factors which can potentially be modified and hence improved.

## (2) Determining the choice of the spatial scale

The selection of spatial scale in this study depended heavily upon data availability. Preliminary data searches made clear that sufficient data on the various risk factors of dementia were generally only publicly available at the Clinical Commissioning Group (CCG), local authority, sub-regional or regional level. A key point of interest for future research is that substantial amounts of National Health Service data are aggregated to the CCG level from constituent GP practices. Should this data be made available from source, it would then be possible to apply this IMDR at a finer spatial resolution and target interventions more specifically. However, given that key responsibilities for the commissioning of healthcare are made at CCG level, (for example; urgent and emergency care, health services, elective hospital services, community care, and mental health (NHSCC, 2018)), the application of the index at this scale does provide the opportunity to act upon dementia prevention activities.

## (3) Sourcing the input variables

This step involved sourcing suitable data to capture each risk factor within the IMDR. The data used to represent the 9 risk factors are listed in Table 2, data were selected for 2013/14, or the closest available year, to allow commonality (in time) between data sources.

| Risk factor | Indicator & time period | Source & link |
|---|---|---|
| *Obesity* | *Recorded prevalence of Obesity (16+) (2013/14)* | Public Health England: http://fingertips.phe.org.uk/dementia |
| *Excessive Alcohol Consumption* | *Modelled prevalence of binge drinkers (16+) (2006-2008)* | NHSIC 2010 (now NHS Digital), as from PHE-Local Health: http://localhealth.org.uk |
| *Education* | *GCSE Achievement (5A\*-C inc. English & Maths) (2013/14)* | Department of Education, as from PHE-Local Health: http://localhealth.org.uk |
| *Hearing Loss* | *Prevalence of hearing loss by CCG area hearing loss of at least 25 dBHL: Estimated prevalence (2014)* | NHS England: https://england.nhs.uk/publication/prevalence-of-hearing-loss-by-local-authority-area-2014-ons-estimates |
| *Smoking* | *Smoking: Recorded prevalence (15+) (2013/14)* | Public Health England: http://fingertips.phe.org.uk/dementia |
| *Depression* | *Depression: Recorded prevalence (18+) (2013/14)* | Public Health England: http://fingertips.phe.org.uk/dementia |
| *Hypertension* | *Hypertension: Recorded prevalence (all ages) (2013/14)* | Public Health England: http://fingertips.phe.org.uk/dementia |
| *Healthy Diet* | *Healthy eating adults: Estimated prevalence (16+) (2006-2008)* | Department of Education as from PHE-Local Health: http://localhealth.org.uk |
| *Diabetes* | *Depression: Recorded prevalence (18+) (2013/14)* | Public Health England: http://fingertips.phe.org.uk/dementia |

*Table 2: Input variables for IMDR and source of data*

**Indicator notes:**

1. Census qualification data consisting of 'education levels' were deemed unsuitable given how a broad range of qualifications are grouped together. Livingston et al (2017) specifically identified educational attainment at secondary level as the risk factor.

2. Modelled data underlying the '*diet*' and '*excessive alcohol consumption*' indicators were created using the Health Survey for England, 2007-2008 (HSCIC, 2007). Originally produced for use at Middle Layer Super Output Area level (2001), CCG level estimates were made available by Public Health England (PHE).

3. Hearing loss prevalence estimates were produced by PHE using estimates from 'Hearing in Adults'. Data were converted from the founding 2013 CCG boundaries to the latest 2017 revision. These were simply accretions of smaller CCGs to a larger body (Newcastle & Gateshead, Manchester) but in the case of the newly named Morecambe Bay and North Cumbria CCGs minor border changes occurred, though the impact is relatively minor. Furthermore, the data were converted to prevalence rate data using ONS 2013/14 mid-year population estimates.

As discussed previously, age has always been deemed as the primary dementia risk factor, however, this may be due to various compounding influences accumulating over the life course and it should be noted that dementia is by no means an inevitable consequence of reaching older age (Livingston et al, 2017). Whilst age must be considered in dementia risk, for the purposes of this index which aims to determine *modifiable* dementia risk, it has not been included. As past literature indicates, age does not contribute significant risk until age 65+, after which risk increases exponentially. This also means that potentially modifiable risk factors have already had a lifetime's impact. Furthermore, difficulties exist in including age ranges, for example one CCG's age structure may heavily skew towards a younger population, while another may skew towards a middle-aged population. In this case, decisions in the weighting of modifiable risk factors would have drastic impacts, because very few young people suffer from hearing loss, or the educational achievement at 16 may impact less on an older population.

Additionally, other variables were identified as key modifiable risk factors in the literature but were omitted from the index for the reasons noted: Social isolation (data unavailable, other than via indices or proxies (see Lucy and Burns 2017)), physical activity (data unavailable at the required scale, although the omission is mitigated somewhat through the inclusion of diet), sleeplessness (the evidence for this risk factor is promising but more confirmatory studies are required) and living near a major road (this is difficult to utilise at the large CCG scale plus further confirmatory studies are required).

## (4) Pre-processing the data

Before generating the IMDR, the data underwent a series of analyses and transformations to ensure their suitability. These included: Correlation analysis, normalisation, checking for polarity and weighting.

A multivariate **correlation analysis** was conducted to assess the relationships between input variables. As reflected in Table 3, most variables are interrelated and statistically significantly as expected, yet generally not overly correlated, and are still clearly measuring different components of dementia risk.

| **Indicator**      | (1)     | (2)      | (3)     | (4)      | (5)     | (6)     | (7)     | (8)   |
|---|---|---|---|---|---|---|---|---|
| **(1)** *Education*   | 1       | -        | -       | -        | -       | -       | -       | -     |
| **(2)** *E. Alcohol.C* | 0.15*   | 1        | -       | -        | -       | -       | -       | -     |
| **(3)** *Healthy Diet*| 0.65**  | 0.43**   | 1       | -        | -       | -       | -       | -     |
| **(4)** *Hear. Loss*  | 0.03    | 0.23**   | 0.22**  | 1        | -       | -       | -       | -     |
| **(5)** *Smoking*     | 0.74**  | 0.25**   | 0.61**  | -0.24**  | 1       | -       | -       | -     |
| **(6)** *Obesity*     | 0.57**  | 0.23**   | 0.75**  | 0.22**   | 0.58**  | 1       | -       | -     |
| **(7)** *Hyperten.*   | 0.28**  | 0.22**   | 0.52**  | 0.85**   | 0.07    | 0.56**  | 1       | -     |
| **(8)** *Diabetes*    | 0.44**  | -0.21**  | 0.53**  | 0.08     | 0.38**  | 0.64**  | 0.44**  | 1     |
| **(9)** *Depression*  | 0.36**  | 0.43**   | 0.51**  | 0.36**   | 0.39**  | 0.44**  | 0.45**  | 0.18* |

*Table 3: Correlation analysis between dementia risk Indicators. (\*) Significant at the 0.05 level (\*\*) Significant at the 0.01 level.*

**Normalisation** of the data was carried out to allow for greater comparability of the various indicators, primarily as this process diminishes the impact of bias on a variable with a large magnitude whilst also appreciating that different variables may run on different numerical scales or ranges. For example, a set of data for 'healthy diet' may range between 60% and 90%, while 'hearing loss' may range between 4% and 8%. Clearly, when these variables are combined into an index, the desire is for each variable to introduce a noteworthy dimension of dementia risk per area – with this driven entirely by the magnitude of the variable relative to the other data points, not by the base range. Without normalisation, the indicator for 'healthy diet' would potentially diminish or eliminate valuable information from that of 'hearing-loss'. Furthermore, as indicators may have different units of measurement, normalisation can assist in producing a common numerical range across all variables. This would be beneficial if considering other risk factors or indicator data (such as distances or counts), however, all variables in this study were available as population rate data.

For simplicity and ease of replication, the min-max normalisation method was selected and carried out using the equation below. This linearly rescales variables between 0 and 1, with 1 being the highest value in the dataset and 0 the lowest, with all the remaining data points falling proportionally in between.

Each individual variable $x$, is normalised, $x_{norm}$ using:

$$x_{norm} = \frac{x - x_{min}}{x_{max} - x_{min}}$$

where $x_{max}$ and $x_{min}$ are the maximum and minimum values in the set of $x$.

Data **polarity** refers to the direction of the variables, in this study some variables are positively correlated with dementia risk (known as positive polarity) and some are negatively correlated (negative polarity). Accounting for polarity is of great importance as, if left unchecked, a high value in one variable may positively impact the measure while a high value in a second variable may negatively impact the measure. To create a simple final index score, the decision was taken that all variables would contribute positive polarity, so each indicator would add to the CCGs dementia risk score.

To highlight the issue in the context of this index, one may consider the variables 'recorded smoking prevalence' which exhibits positive polarity of dementia risk (thus a higher value in this variable may be associated with increased dementia risk), and 'GCSE achievement' which displays negative polarity (thus a higher value in this variable may be associated with decreased dementia risk). Combining these two variables forces the datasets to work in opposite directions, however, if one reverses the polarity of 'GCSE achievement' it would then show the percentage of people who did *not* achieve 5 A\*-C per CCG, thus allowing each variable to positively contribute to the index such that the higher the score of the IMBR the higher the overall dementia risk.

Two variables representing the risk factors of 'education' and 'diet' were identified as having negative polarity. Consequently, these were reversed by subtracting the normalised values from 1. This consequently derived two new variables: [1] Education – those NOT achieving 5 A*-C inc. English & Maths and [2] Diet – those not deemed to be healthy eating adults.

**Weighting** individual component variables is generally an optional step in the generation of any composite index. This process is employed in the IMDR due to the strong evidence presented within the literature highlighting that different risk factors do hold differing overall contributions to dementia risk. Table 4 presents the weightings put forward for this index. These are influenced by the work of Livingston et al (2017) with some adaptions made based on other literature observations, namely the work of Schwarzinger et al (2018) when investigating the impact of alcohol consumption. The final weightings proposed for this index gave slightly more power to lifestyle factors and education, largely due to evidence that these have led to a reduction in population dementia incidence (Norton et al, 2014; Wu et al, 2017; Livingston et al, 2017).

| Risk factor | Weighting in Livingston et al (2017) (%) | Final Adjusted weighting (%) |
|---|---|---|
| Obesity | 2% | 6% |
| E. Alcohol C. | 9% | 21% |
| Education | 8% | 21% |
| Hearing Loss | 9% | 18% |
| Smoking | 5% | 12% |
| Depression | 4% | 8% |
| Hypertension | 2% | 4% |
| Healthy Diet | ($^1$)3% | 8% |
| Diabetes | 1% | 2% |
| TOTAL | ($^2$)41% | 100% |

**Table 4.** *Relative Weighting of Risk Indicators*

*($^1$) Healthy diet replaced physical activity and is weighted accordingly ($^2$) The Livingston PAF model included 65% potentially unmodifiable risk factors (age and genetic factors), social isolation at 3% was not included while E. Alcohol C. was included within the index in this research.*

### (5) Generating index scores

In this research, both a weighted and unweighted version of the IMDR were created.

In order to generate the final unweighted index score per CCG, the normalised values were summed (per CCG) and divided by the total number of risk factors (9), thus creating an average across the 9 risk factors per areal unit. This mirrors the process detailed in Lucy and Burns (2017). The following equation summaries the process for the unweighted index, where $IMDR_j$ = risk index for CCG $j$ and $V_i$ = the total indicator $i$.

$$IMDR_j = \sum_{i=1}^{9} \frac{(V_i)}{9}$$

For the weighted index, the same process was followed except the variables were assigned their respective weights (as presented in Table 4). The following equation summaries the process for the weighted index, where $IMDR_j$ = risk index for CCG $j$ and $V_i$ = the total of indicator $i$ and $W_i$ = weighting of $V_i$.

$$IMDR_j = \sum_{i=1}^{9} \frac{(V_i * W_i)}{9}$$

### (6) Visualisation

To **visualise** the spatial patterns between English CCGs, the weighted and unweighted indices were integrated into a geographic information system (MapInfo Pro) with a geo-referenced map of English CCG boundaries downloaded from the open geography portal website (ONS, 2017). Using choropleth mapping, areas with higher and lower IMDR scores can be distinguished and a description of the geographic variations observed.

## Index results

Figure 4 presents both the weighted (top) and unweighted (bottom) index in spatial form, presented side-by-side for ease of comparison. On visual inspection, there appears to be little difference in the two measures, with only a very select group of CCGs observing any change in value. In the unweighted index, a small number of CCGs in the south gain an increase in score, whilst in the north, Newcastle and Gateshead CCG observes a decrease in score, something which is also mirrored in five rural CCGs around North Yorkshire. The use of standard deviation to determine class intervals clearly visualises the disparity between London and the rest of the country, with the lowest two ranges almost entirely in and around the London vicinity with the exception of Rushcliffe CCG in Nottinghamshire.

Consistent with the literature review, a clear north-south health divide can be observed with regard to the IMDR (Curtis, 1995; Gould and Jones, 1996; Sloggest and Joshi,1994; Hacking et al, 2011; Buchen et al, 2017), although in this case in terms of potentially modifiable dementia risk. A steady transition in risk score from south to north is observed, with several major areas of risk beginning around Birmingham (but not the city itself). London stands out in terms of the high number of low scoring CCGs; all 32 constituent London CCGs fall into the lowest risk map categories.

Generally, areas of low risk appear to emanate out from London with risk increasing as distance from the capital city increases. The size of certain CCG's, such as those in the South-West of England, somewhat dominate this emanation pattern from London in the South of England, however, the spatial trend is consistent for a number of CCGs in the south. Evident in Great Yarmouth and Waveny, Thanet, Portsmouth, Southampton, and Isle of Wight CCGs, all of which display some of the highest IMDR scores in the South of England.

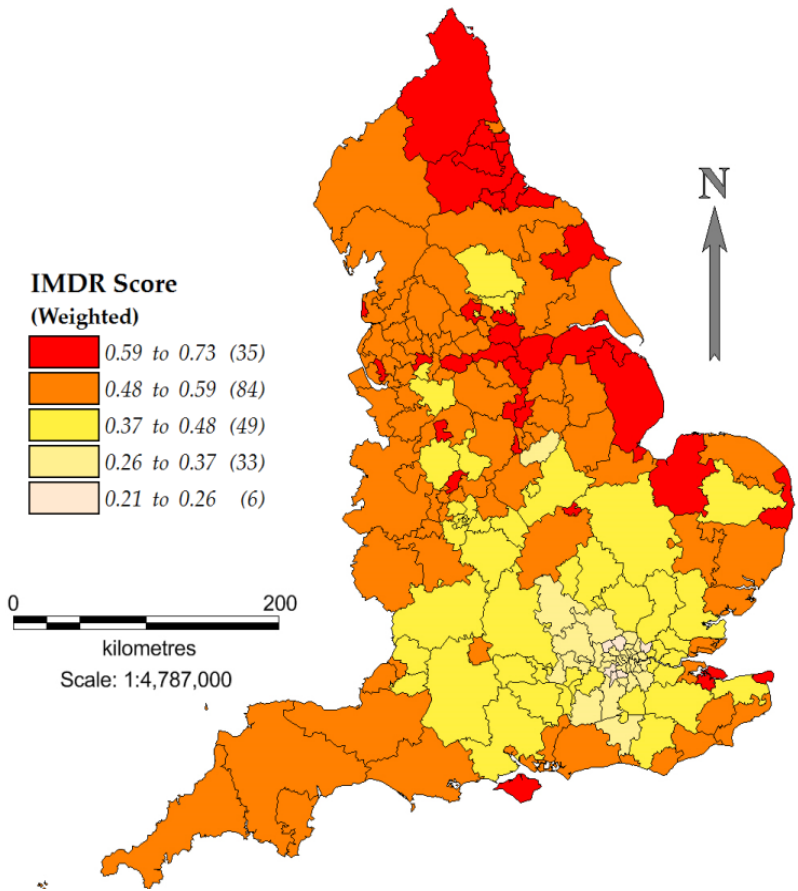
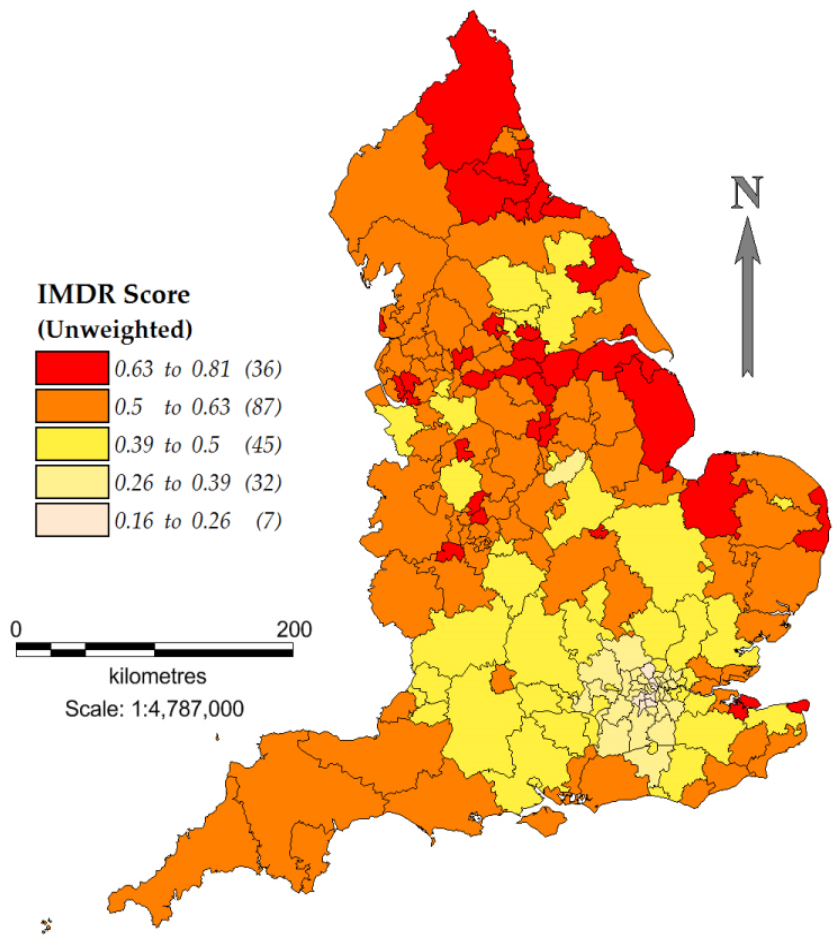

Figure 4: English Clinical Commissioning Groups (CCGs) by Index of Modifiable Dementia Risk (IMDR) score with (top) and without (bottom) weighting.

In the North of England, there are two concentrated areas of risk. Firstly, the CCGs surrounding the Newcastle and Gateshead show one of the highest concentrations of IMDR scores. The second concentration spans a variety of Northern cities, towns, and rural CCGs stretching from Humberside to Merseyside and includes the following CCGs; North and Northeast Lincolnshire, Hull, Doncaster, Barnsley, Wakefield, Leeds South and East, Rotherham, Mansfield and Ashfield, Doncaster, Manchester, Salford, and Liverpool. Contrasting this pattern are Leeds (North) and Harrogate and Rural District CCGs which buck the northern trend and show lower IMDR scores.

Focusing on the CCGs at either extreme of the IMDR scale reveals a continuity in the overall north-south divide and London versus the rest of England pattern. The IMDR scored Blackpool, Sunderland, and Barnsley CCGs as having the highest potentially modifiable dementia risk, conversely Harrow, Redbridge, and Barnsley CCGs scored lowest.

Figures 5 and 6 display the factors which contributed to each of these CCGs IMDR scores. In Figure 5, the areas of highest risk as ranked by the IMDR share the same pattern in terms of the elements which contributed to the high-risk score.

As suggested previously, all variables within the final IMDR add to an areas potentially modifiable dementia risk score. The largest contributing factors for each high scoring CCG were: excessive alcohol consumption, as measured by the estimates of binge drinking prevalence; education, which used GCSE achievement (5 grade A*-C) at 16; and hearing loss, assessed with estimated data of prevalence. In the IMDR, education and hearing loss were weighted as the most important factors (21% & 18% respectively) following the risk-scores obtained in the PAF model of Livingston et al (2017), and excessive alcohol consumption was weighted highly (21%) due to the large-scale study of over 30 million French patients as discussed by Schwarzinger et al (2018). As these factors account for much of the risk of dementia, it is understandable that they are chief contributors to the overall IMDR score. Across 8 of the 9 risk factors, Blackpool CCG shared a consistent score with other low scoring CCGs, however, smoking prevalence (recorded at GP level within each CCG) was high enough to allow it to receive the highest overall IMDR score.

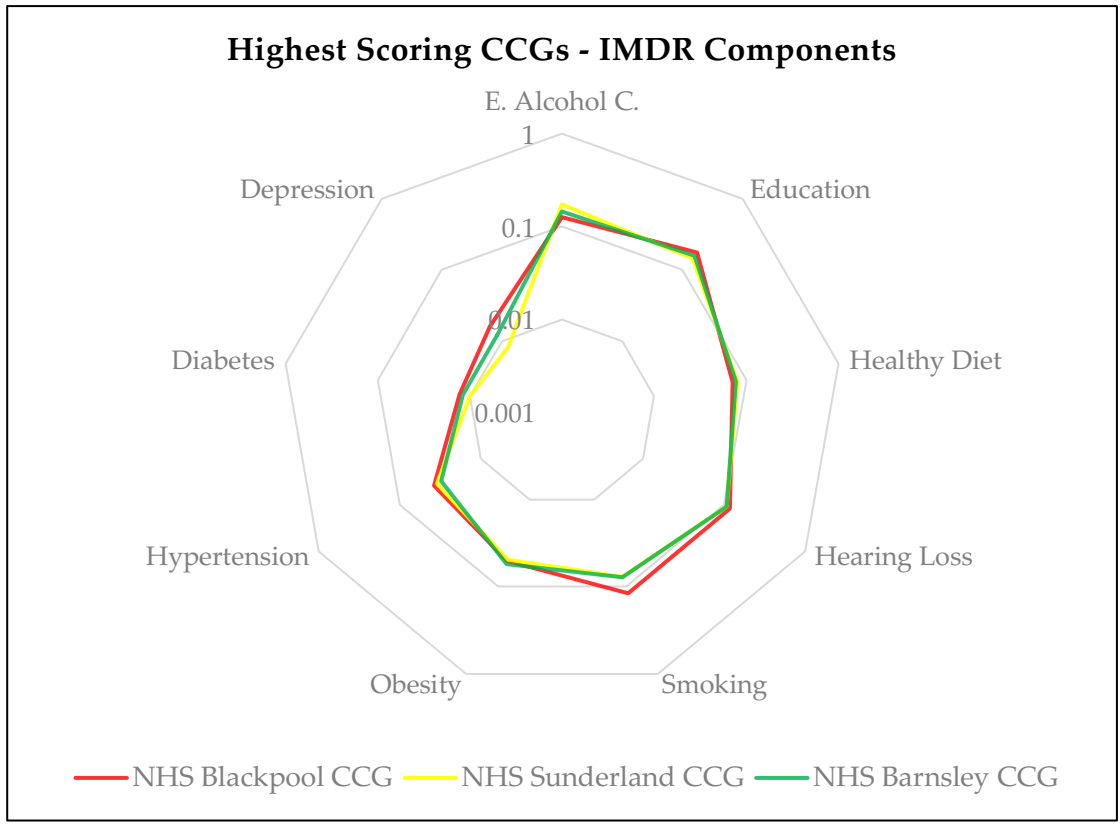

*Figure 4: Radar Plot showing constituent elements of top CCG IMDR scores*

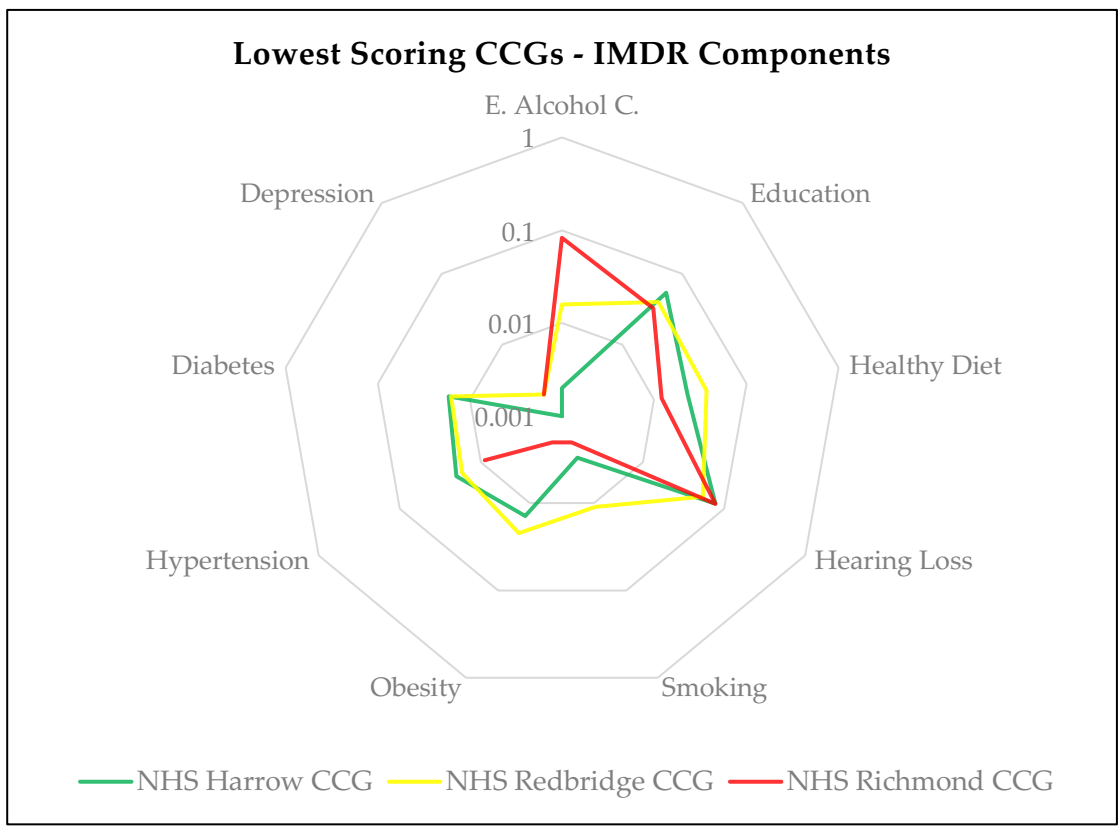

*Figure 5: Radar Plot showing constituent elements of bottom CCG IMDR scores*

Figure 6 shows much more variation between the lowest scoring CCGs, however, because the scores of each factor remain so low, the overall contribution of any of the factors is limited. Hearing loss contributes the most in each of these low-risk areas yet remains below the magnitudes observed in Figure 5. Interestingly, despite having an entirely insignificant prevalence of diabetes and generally much lower scores, Richmond CCG would have been the lowest ranked CCG if not for the disproportionate contribution from excessive alcohol consumption. This shows the sensitivity of the index to change in various risk factors, which may not be entirely useful when comparing such large and heterogenous areas.

Table 5 presents the population count, Index of Deprivation (IMD) score, recorded and estimated dementia prevalence and the average age of the top and bottom performing IMDR scoring CCGs. The IMDR maintains a generally consistent trend with average age and estimated dementia prevalence, and a largely positive association with recorded dementia prevalence, in that increases in IMDR are matched by subsequent increases in these factors. The relationship between the IMD and IMDR appears consistent for the top 3 northern CCGs, however, is disrupted in the case of Richmond, Redbridge, and Harrow CCGs.

| NHS CCG | IMDR (Weighted) | *Population (Total)* | IMD (2015) | Dementia Prevalence: *Recorded (left) Estimated (right)* | | Average age |
|---|---|---|---|---|---|---|
| *Blackpool* | 0.7300 | *139,195* | 42.000 | 5.1 | 78.5 | 42.00 |
| *Sunderland* | 0.6900 | *277,962* | 29.730 | 4.3 | 72.1 | 41.00 |
| *Barnsley* | 0.6800 | *241,218* | 29.570 | 4.2 | 70.6 | 41.00 |
| *Richmond* | 0.2200 | *195,846* | 10.040 | 4.2 | 67.7 | 39.00 |
| *Redbridge* | 0.2200 | *299,249* | 20.240 | 4.5 | 64.7 | 36.00 |
| *Harrow* | 0.2100 | *248,752* | 14.300 | 4.2 | 65.1 | 38.00 |

*Table 5: Top and bottom CCG statistics*

# Validation, corroboration and further analysis

In a bid to further understand the index patterns and at the same time gauge the overall robustness of the index, the results were contrasted with data on deprivation (IMD), recorded dementia prevalence and average age. The spatial patterns observed in Figure 4 (weight and unweighted) are used as the basis of comparison between the IMDR and each of the aforementioned factors (shown in Figure 6).

Recall that the IMDR tends to reflect lower scores in/around London with scores increasing as distance from the capital increase. This leads to high scores in northern England (Humberside to Merseyside) and lower scores in southern England, with the lowest scores within London's 32 constituent CCGs.

When reviewing Figure 6, similar patterns are apparent with respect to deprivation. Lower IMD scores surround and emerge out of London, however, the majority of CCGs in London score highly in this measure of deprivation. On a wider scale, very similar north-south patterns are clear, with northern England experiencing greater deprivation.

In terms of recorded dementia prevalence, the pattern in London is the direct opposite to that of the IMDR. There are increased levels of dementia prevenance in London's constituent CCGs. The north-south divide is similar to the IMDR however, although the overall pattern is diminished with no clear corridor from Humberside to Merseyside.

Finally, when exploring average age, the patterns reflected in the IMDR transfer to this variable in that London's CCG's have a generally low average age and this statistic increases with distance from the capital (more broadly). For this reason, similar north-south patterns also occur here.

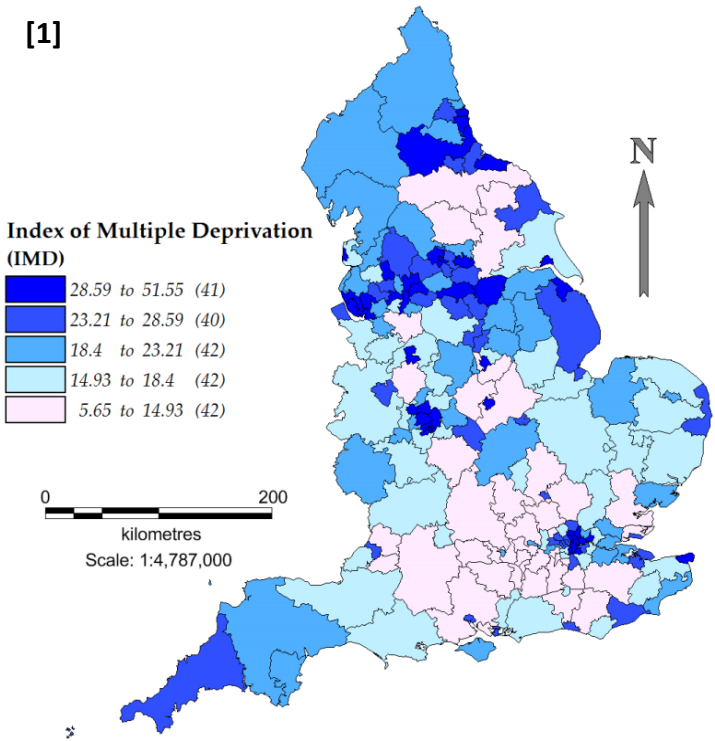
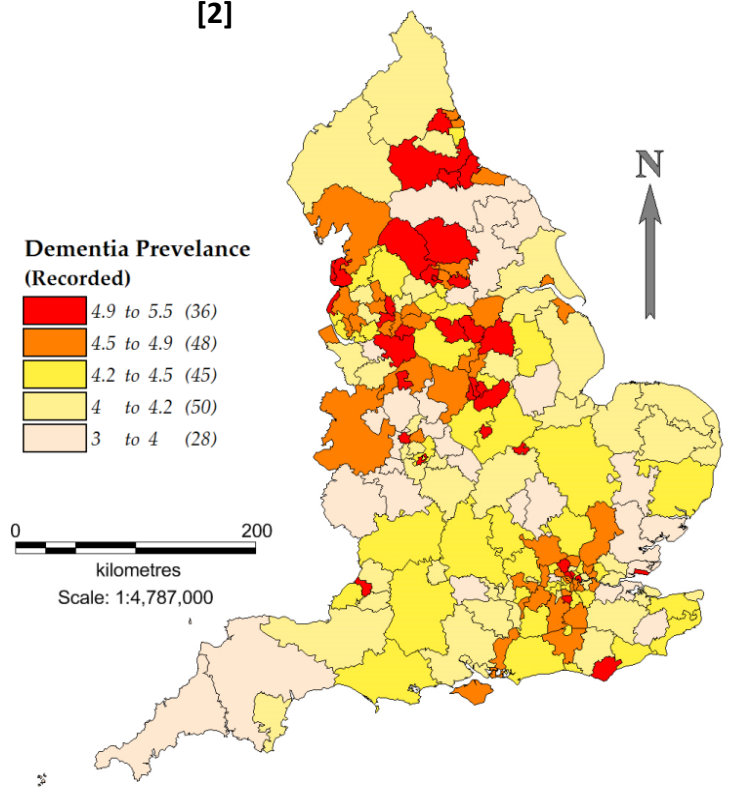
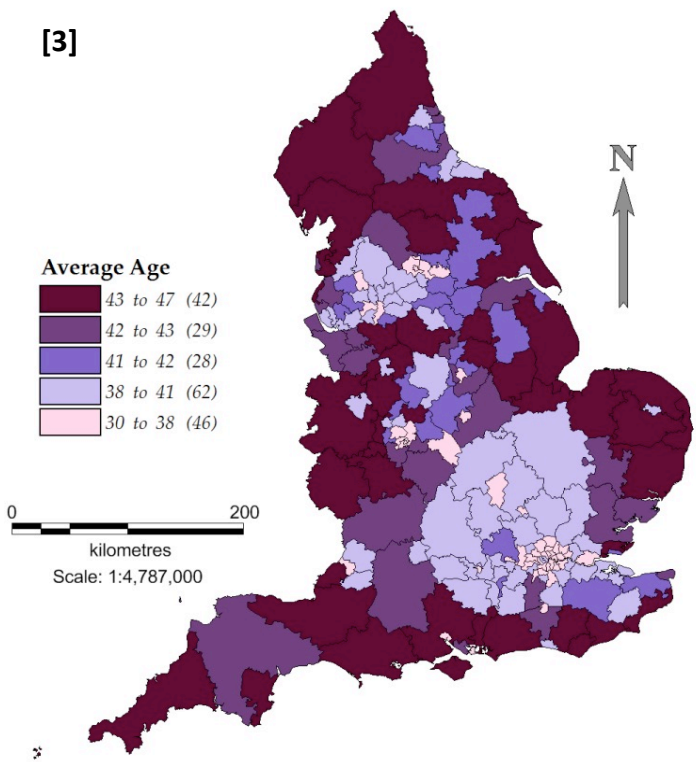

*Figure 6: English Clinical Commissioning Groups (CCGs) by [1] Index of Multiple Deprivation (2015), [2] Recorded Dementia Prevalence and [3] Average Age. All data referenced previously.*

# Additional analysis

As a final investigative step, emergent patterns between both the weighted and unweighted IMDR and the IMD, recorded and estimated dementia prevalence, and average age will be further assessed. The statistical analysis attempts to validate the weightings used in the production of the weighted IMDR to endorse it as a more accurate areal dementia risk measure.

Figure 7 displays a histogram of all CCGs ranked by weighted IMDR score. It can be observed that the data are rather skewed, yet generally normally distributed around IMDR scores of 0.5. Relating this to Figure 4, visually it is possible to identify the group of values which skews this distribution – these are centred on the 32 London CCGs. The IMDR, as applied at this large scale, appears to break down at a national level somewhat due to the uniqueness of London. As such, the following overall Pearson correlations are conducted for all 207 CCGs and repeated excluding the 32 London CCGs (see Tables 6 and 7).

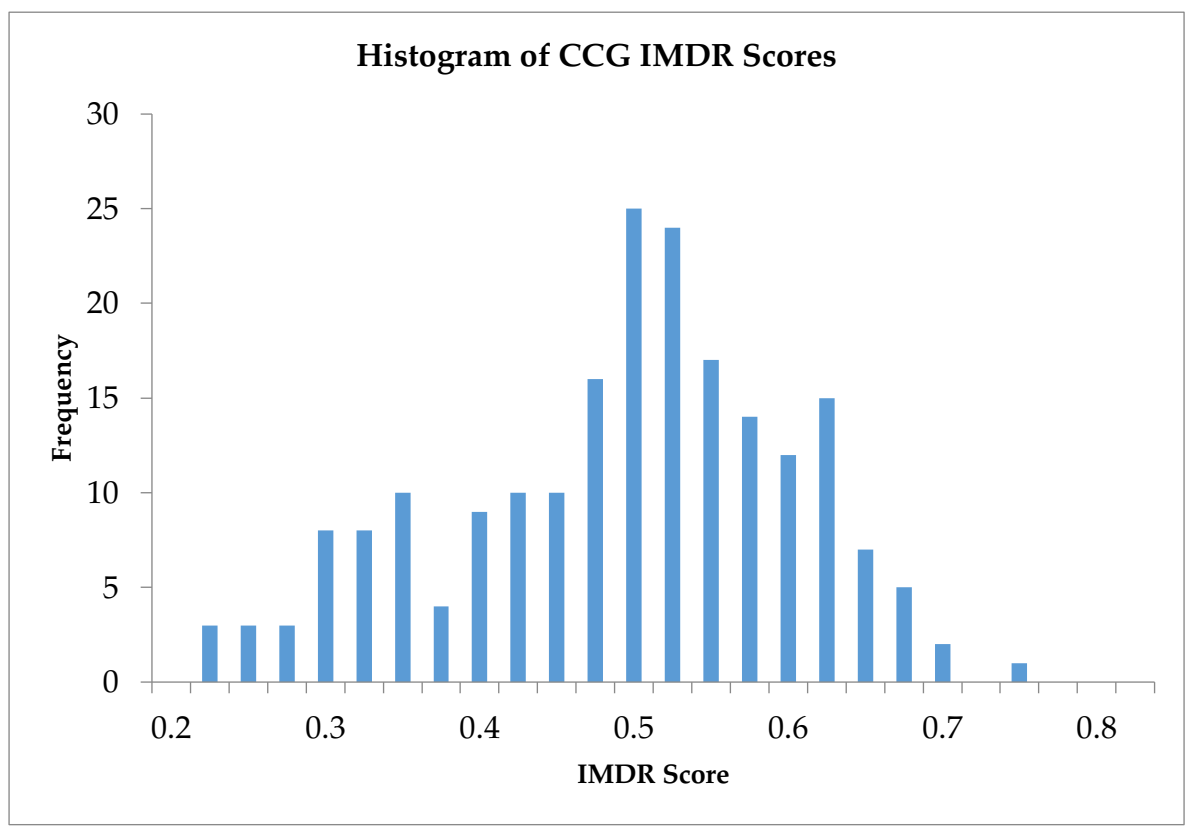

*Figure 6: CCGs by weighted IMDR Score*

| Measure | (1) | (2) | (3) | (4) | (5) | (6) |
|---|---|---|---|---|---|---|
| **(1)** *IMDR (W)* | **1** | - | - | - | - | - |
| **(2)** *IMDR (Uw)* | .960** | **1** | - | - | - | - |
| **(3)** *IMD (2015)* | .388** | .413** | **1** | - | - | - |
| **(4)** *Dem.Rec.Prev.* | .021 | .004 | .316** | **1** | - | - |
| **(5)** *Dem.Est,Prev.* | .113 | .095 | 0.474** | .856** | **1** | - |
| **(6)** *Avg. Age* | .447** | .435** | -.473** | -.250** | -.332** | **1** |

*Table 6: IMDR and Validation Measures Correlation analysis (for all 207 English CCGs)*
*(\*) Significant at the 0.05 level (\*\*) Significant at the 0.01 level (df=204)*

Table 6 presents the results of the Pearson Correlations between the IMDR and the aforementioned comparators. Unsurprisingly, the weighted IMDR and unweighted IMDR are extremely correlated (r=.960 p=.001) as they contain the same data and the weightings only slightly adjust certain risk factors. Interestingly, the unweighted IMDR has a stronger correlation with the IMD (r=.388, p=.001) compared to the weighted IMDR (r=.413, p=.001), with both displaying statistically significant weak-moderate correlations. This pattern is reversed for the relationships of the weighted and unweighted IMDR with both predicated and estimated dementia prevalence, however, all four are statistically insignificant.

| Measure | (1) | (2) | (3) | (4) | (5) | (6) |
|---|---|---|---|---|---|---|
| **(1)** *IMDR (W)* | **1** | - | - | - | - | - |
| **(2)** *IMDR (Uw)* | .945** | **1** | - | - | - | - |
| **(3)** *IMD (2015)* | .591** | .613** | **1** | - | - | - |
| **(4)** *DemRec.Prev.* | .114* | .110 | .324** | **1** | - | - |
| **(5)** *Est,Prev.* | .303** | .293** | 0.455** | .883** | **1** | - |
| **(6)** *Avg. Age* | .209** | .219** | -.474** | -.234** | -.288** | **1** |

*Table 7: IMDR and Validation Measures Correlation analysis (English CCGs, excluding 32 London CCGs)*
*(\*) Significant at the 0.05 level (\*\*) Significant at the 0.01 level (df=172)*

Table 7 shows the same Person Correlations but with the 32 CCGs which comprise London excluded (due to the lowly skewed IMDR scores and age differences displayed previously). This change presents significantly stronger and correlations for both the weighted and unweighted IMDR. The previous correlation patterns are maintained and are considerably strengthened; whereby the unweighted IMDR, compared to the weighted IMDR, displays a slightly higher correlation with the IMD (r=.591, p=.001) and lower correlations with recorded prevalence (r=.110, sig=.009) and estimated dementia prevalence (r=.293, p=.001), whereas the weighted IMDR, in comparison to the unweighted IMDR, displays a slightly lower correlation with the IMD (r=.591, p=.001) yet higher correlations with both the recorded (r=.114, p=.001) and estimated (r=.303, p=.001) dementia prevalence.

Finally, Table 6 shows correlations between the weighted IMDR and age (r=.447, p=.001) – this reflects the spatial patterns observed previously, and is understandably weakened in Table 7 (r=.209 p=.001) with the exclusion of London CCGs. As per Table 6, age also displayed statistically significant weak-moderate negative correlations with the IMD (r= -.473 p=0.01) recorded (r= -.250, p=.001) and estimated (r= -.332, p=.001) dementia prevalence.

As a final step, due to the spatial similarities between Figures 4 and 6 [3], the impact of age as a confounding variable was assessed by running a partial correlation between the IMDR and the other validation measures, controlling for the average age of each CCG. This resulted in an increased strength of the previously observed correlations, as evident in Table 8. Looking only at the weighted IMDR, the strength of the correlation between this and the IMD is surprising strong at r=7.60 (p=.001), which compares favourably with the relationships observed in Hudson (2005) and Letellier et al (2017). Additionally, the NHS dementia estimates displayed a statistically significant (although weak) correlation with this researches measure of modifiable dementia risk when controlling for age r=.310 (sig=.001), and also recorded prevalence displays a weaker yet statistically relevant correlation when controlling for age r=.153 (sig=0.05) – this adds lends credence to the IMDR.

| **Measure** | **(1)** | **(2)** | **(3)** | **(4)** | **(5)** | **(6)** |
|---|---|---|---|---|---|---|
| **(1)** *IMDR (W)* | 1 | - | - | - | - | - |
| **(2)** *IMDR (Uw)* | .950** | 1 | - | - | - | - |
| **(3)** *IMD (2015)* | .760** | .780** | 1 | - | - | - |
| **(4)** *DemRec.Prev.* | .153* | .121 | .231** | 1 | - | - |
| **(5)** *Est,Prev.* | .310** | .282** | 0.381** | .847** | 1 | - |

*Table 8: Partial correlation analysis IMDR and Validation Measures, controlling for Average age per CCG (*) Significant at the 0.05 level (**) Significant at the 0.01 level*

# Conclusion

This research sought to spatially and statistically examine patterns of modifiable dementia risk factors through the creation of a spatial index, named the Index of Modifiable Dementia Risk (IMDR), and by evidencing the potential usefulness of such an index by applying it to English Clinical Commissioning Groups (CCGs). Optimistically, such an index could be recreated by healthcare planners or local government for use at a smaller geographic scale, adding or removing relevant risk factors within the specific context of the area or as new evidence on dementia risk becomes available.

Overall, this paper detailed the design and utility of an index to spatially quantify potentially modifiable dementia risk, defined as those risks which are conceivably modifiable via changes in lifestyle, such as diet and exercise; active medical intervention, for instance treatment of hypertension and management of hearing loss; or population level initiates, for example better education and anti-smoking campaigns. Social, economic and health disparities between the north and south of England have long been identified (Gould and Jones, 1996) yet persist (Buchan et al, 2017) and were found in this research. As knowledge of risk factors, data availability and research methods improve such an index can be continually renewed and updated. Nonetheless, society is already at a point in time where changes can be made and individual dementia risk modified (Livingston et al, 2017). It is also known that a social gradient to health exists (Marmot and Bell, 2012) and while life expectancy is increasingly globally, life expectancy free from disease is severely lagging behind, particularly for the most deprived in society (GBD, 2016). This research has shown that there is certainly a role for spatial analytics within mental health and this gap in the health geography literature is severely underserved. Information about relevant areas of concern, for example, is crucial in both healthcare planning, deployment of resources and intervention opportunities. Individual behaviours, outcomes and lifestyle choices remain key areas of policy concern, and as evidence has shown, there remain many possibilities in improving public health as a whole and reducing risk of dementia.